\documentstyle[psfig]{mn}

\newif\ifAMStwofonts
\AMStwofontstrue

\def\spose#1{\hbox to 0pt{#1\hss}}
\newcommand\lsim{\mathrel{\spose{\lower 3pt\hbox{$\mathchar"218$}}
     \raise 2.0pt\hbox{$\mathchar"13C$}}}
\newcommand\gsim{\mathrel{\spose{\lower 3pt\hbox{$\mathchar"218$}}
     \raise 2.0pt\hbox{$\mathchar"13E$}}}

\title[Detecting GWs coincident with GRBs] {A method for detecting
      gravitational waves coincident with gamma ray bursts}

\author[M. T. Murphy, J. K. Webb and I. S. Heng]
       {M. T. Murphy$^1$\thanks{E-mail: mim@bat.phys.unsw.edu.au (MTM)}, J. K. Webb$^1$ and I. S. Heng$^2$ \\
     $^1$School of Physics, The University of New South Wales, Sydney 2052, Australia.\\
     $^2$Department of Physics, University of Western Australia, Nedlands, WA 6907, Australia.\\}

\date{Accepted 2000 March 9.
      Received 2000 February 2;
      in original form 1999 May 25}

\pagerange{\pageref{firstpage}--\pageref{lastpage}}
\pubyear{1999}

\begin{document}

\maketitle

\label{firstpage}

\begin{abstract}
The mechanism for gamma ray bursters and the detection of gravitational
waves (GWs) are two outstanding problems facing modern physics. Many models
of gamma ray bursters predict copious GW emission, so the assumption of an
association between GWs and GRBs may be testable with existing bar GW
detector data. We consider Weber bar data streams in the vicinity of known
GRB times and present calculations of the expected signal after co-addition
of 1000 GW/GRBs that have been shifted to a common zero time. Our
calculations are based on assumptions concerning the GW spectrum and the
redshift distribution of GW/GRB sources which are consistent with current
GW/GRB models. We discuss further possibilities of GW detection associated
with GRBs in light of future bar detector improvements and suggest that
co-addition of data from several improved bar detectors may result in
detection of GWs (if the GW/GRB assumption is correct) on a time scale
comparable with the LIGO projects.
\end{abstract}


\section{Introduction}
The attempts to detect gravitational radiation began with the work of Weber
(1960). The existence of gravitational waves has been verified indirectly
by observations of the decay in orbital period of the pulsar PSR 1913+16
(Hulse \& Taylor 1975), for which the 1993 Nobel Prize in Physics was
awarded. However, direct detection continues to prove elusive even though
some authors believed it to be possible before the turn of the century
(eg. Thorne 1992). Detection remains difficult due to the nature of the
coupling between energy and spacetime. Although a large amount of energy
might be stored in gravitational waves (GWs), the corresponding amplitude
of the waves, measured in dimensionless strain, $h$, is exceedingly small
due to the very small magnitude of the proportionality constant in
Einstein's equations: a large stress-energy produces a very small metric
perturbation.

The most sensitive detectors currently in operation are of the Weber bar
variety. The main detectors are maintained at the University of Western
Australia (eg. Heng et al. 1996), Louisiana State University (eg. Mauceli
et al. 1996), the University of Rome (eg. Astone et al. 1993), Legnaro
National Laboratories (eg. Cerdonio et al. 1995) and Laboratori Nazionali
di Frascati (eg. Coccia et al. 1995). These detectors have a very low
bandwidth as they rely on detection of the resonance in the bar (or sphere)
after a burst of GWs. Near the resonance frequency, bar detectors are
capable of reaching sensitivities of $\sim 10^{-19}$--$10^{-18}$ (eg.
Tobar et al. 1995) and future sensitivities are expected to reach $\sim
10^{-20}$ (eg. Tobar et al. 1998). Spurious, non-thermal excitations of
resonant bars typically number $\sim 40$ per day (Heng et al. 1996).  Thus,
detection of a real, individual GW burst will require coincidence with
detections by other instruments around the world. No convincing
coincidences have been recorded to date.

The LIGO/VIRGO network is expected to detect gravitational waves (Finn \&
Chernoff 1993; Cutler \& Flanagan 1994; Thorne 1994; Flanagan \& Hughes
1998), if not in the initial stages when the sensitivity of LIGO is $\sim
10^{-21}$, then in the advanced stage when it is $\sim 3 \times
10^{-23}$. This will mark the beginning of GW astronomy, promising views of
neutron star (NS) and black hole (BH) coalescence and collisions. It will
also yield detailed information about the dynamics of pulsar systems and
may even allow us to study inflation and the Big Bang itself more directly
and in much more detail than other astrophysical techniques allow.

Another problem that has confronted science for decades has been the origin
of gamma ray bursts (GRBs). Many theories for the emission of GRBs centre
on a mechanism involving large mass quadrupole moments.  Hence a wide
variety of GRB models predict large GW luminosities. For example, NS
mergers are good candidates for GRBs (eg. Kochanek \& Piran 1993;
Dokuchaev, Eroshenko \& Ozernoy 1998) and are expected to result in $\sim
10^{52}{\rm ~ergs}$ in GWs (Centrella \& McMillan 1993; Ruffert \& Janka
1998).  Such energies result in detectable GW signals if the NSs are in the
Galaxy but if the source is cosmological both modern bar detectors and the
advanced LIGO system cannot detect such events.

It is this expected association between the two phenomena that has prompted
the present investigation. With the expectation that GW bursts are
temporally correlated with GRBs, we may co-add existing GW data surrounding
measured GRB events. This would result in a composite GW data stream with
all GRB events shifted to a common `GRB time'. We explore this idea below
and present a method offering up to a factor of $\sim 30$ increase in
signal-to-noise ratio (SNR) with the co-addition of 1000 existing GW data
streams surrounding measured GRB times. Throughout the work, we assume that
all the measured GRBs are associated with GW emission.

In this article we calculate the expected signal from a typical bar
detector assuming that the bar is noiseless. Our aim is to show directly,
with some generality, whether available data from GRB and GW detectors can
be used to detect GWs before the first LIGO/VIRGO detection. Such a
detection would also be confirmation of an association between the GRB
mechanism and GW emission.

Our calculations are based on various assumptions concerning source
distribution in spacetime, the nature of the source and the corresponding
GW emission, all in line with current models. We compare our calculations
with the noise floor of the bar and decide on the viability of such a
method for detection using bar detectors. We also discuss the viability of
detection by the LIGO system.

The paper is organized as follows. In Section 2 we provide the formalism
for our calculations, discussing the assumptions concerning the GW source
and detector. Various models are considered in Section 3 that are
consistent with the assumption that GRBs are associated with GW
emission. From this discussion we extract features such as the redshift
distribution of sources and the spectrum of the source. In Section 4 we
compute the signal seen at the detector for a single GW/GRB. We also
simulate the co-addition of bar data with the assumption of a common
temporal zero point and we discuss the results for the various models in
Section 4. In Section 5 we discuss the relevance of our result for the
LIGO/VIRGO experiments and provide our conclusions.

\section{Formalism}
Our aim is to calculate the signal seen at the output of a GW bar detector,
with a single resonance band, due to a single GW burst. However, we keep in
mind that we ultimately wish to simulate the composite effect due to many
bursts (shifted to a common burst time); we wish to obtain a Monte Carlo
estimate of the signal at the bar output after data processing. We
therefore assume that there exists a distribution of GW/GRB sources in
spacetime according to the normalized redshift distribution,
$\phi(z)dz$. That is, we require that $\int \phi(z)dz = 1$. To calculate
the effect on the detector we need to find the GW spectrum at the bar due
to a source under this distribution. Therefore, let us assume that the
frequency of GWs at the source is described by the spectral profile
$f(\nu)d\nu$. Again, $\int f(\nu)d\nu = 1$. Also, since we are considering
the spectrum of the source, we may calibrate the GW flux scale by assuming
that the average total energy emitted in GWs by the source is $F$. We
explain our reasoning below.

\subsection{GW emission and propagation}
Our first assumption is of a distribution of mean total GW fluxes at the
detector. The analysis is based on the assumption that GRBs are associated
with GW emission. We further assume that the {\it observed} luminosity
distribution of GWs in the detector frame follows that of GRBs. This
assumption is valid only if the mechanism for gamma ray and GW emission is
closely physically related. We make this assumption on the basis that our
results may then be used as constraints for such a subset of GW/GRB
theories and because of the lack of other reasonable models for the
luminosity distribution of GW sources under the GW/GRB assumption. The
knowledge of whether this relationship holds in reality would be invaluable
for theories of gamma ray bursters.

It is common to define the Earth frame luminosity distribution of GRBs in
terms of the number of bursts with peak gamma ray flux greater than a given
peak flux, $P$, as $N_{\rm {\sc grb}}(>P)$. The fact that this distribution
strays from a $-3/2$ power law at low $P$ is an indication of the
cosmological origin of GRBs. This has been confirmed by several optical
transient observations and redshift determinations (eg. Metzger et
al. 1997; van Paradijs et al. 1997; Djorgovski et al. 1998; Kulkarni et
al. 1998). $P$ is usually defined in terms of a given Earth frame energy
band and the integration time over which the flux was allowed to
accumulate. We use the distribution measured by the BATSE instrument --
supplied in the BATSE 4Br catalogue (Paciesas et al. 1999) -- choosing the
$50$--$300{\rm ~keV}$ energy band and the $64{\rm ~ms}$ integration time. In
our calculations, it is convenient to find the cumulative distribution of
total GW fluxes. We define this as
\begin{equation}
\label{eq:lumdist}
n_{\rm {\sc gw}}[<P(S^{\prime})] = 1 - \frac{N_{\rm {\sc grb}}(>P)}{N_{\rm
{\sc grb}}}
\end{equation}
where $N_{\rm {\sc grb}}$ is the total number of GRBs in the distribution and
$P(S^{\prime})$ is defined as
\begin{equation}
\label{eq:Sdef}
P(S^{\prime}) = \frac{\bar{P}}{\bar{S^{\prime}}}\,S^{\prime}\, .
\end{equation}
Here, $\bar{P}$ is the average value of the peak flux in gamma rays and
$\bar{S^{\prime}}$ is the mean flux in GWs at the detector.

$\bar{P}$ is easily computed from the data in the BATSE 4Br
catalogue. We obtain $\bar{S^{\prime}}$ by evolving $F$ with cosmology
from the source to the detector. This is particularly simple in the
case of a flat universe in which the cosmological constant is
vanishingly small (Misner, Thorne \& Wheeler 1973, p. 783),
\begin{equation}
\label{eq:MTW-S}
\bar{S^{\prime}}(F,z_{\rm m}) = \frac{F}{4\pi (1+z_{\rm m})^2 {\cal R}(z_{\rm m})^2}
\end{equation}
where the distance, ${\cal R}(z_{\rm m})$ is defined by
\begin{equation}
\label{eq:MTW-r}
{\cal R}(z_{\rm m}) = \frac{2 + 2z_{\rm m} - 2\sqrt{1+z_{\rm m}}}{H_0(1+z_{\rm m})}\, .
\end{equation}
Here, $H_0$ is the Hubble constant and $z_{\rm m}$ is the mean of the redshift
distribution ($c$ has been set to unity in Eqs. \ref{eq:MTW-S} and
\ref{eq:MTW-r}). Throughout this work, we use $H_0 = 75{\rm
~kms^{-1}Mpc^{-1}}$. So, by choosing a model of the source, we can estimate
$F$ and normalize the flux scale of $n_{\rm {\sc gw}}(<P(S^{\prime}))$ to
the units of GW flux using Eq. \ref{eq:Sdef}.

Since we are interested in the frequency space at the detector, we may
convert the assumed redshift distribution, $\phi(z)dz$, to a GW frequency
distribution at the detector,
$\phi_{\nu}^{\prime}(\nu^\prime)d\nu^{\prime}$, using the simple
transformation, $1+z = \nu/\nu^{\prime}$. This leads directly to
\begin{equation}
\label{eq:ztof}
\phi_{\nu}^{\prime}(\nu^{\prime}) = (\nu^{\prime})^{-2}\phi\left(\frac{\nu}{\nu^{\prime}} - 1\right)
\end{equation}
provided that we have a value for $\nu$, the frequency of GW at the source
at redshift $z$. We interpret $\nu$ as being the mean frequency at the
source under the distribution $f(\nu)d\nu$. It is also convenient to define
the cumulative frequency distribution at the detector,
\begin{equation}
\label{eq:cumnu}
\Phi_{\nu}^{\prime}(\nu^{\prime}) = \int_{0}^{\nu^{\prime}} \phi_{\nu}^{\prime}(\xi)d\xi\, .
\end{equation}

Since we ultimately wish to co-add the signals from many GWs bursts
associated with GRBs, we may choose a luminosity from $n_{\rm {\sc gw}}(<
S^{\prime})$ by randomly selecting a value for $n_{\rm {\sc gw}}$ and
finding the corresponding value of $S^{\prime}$. Similarly, we use
Eq. \ref{eq:cumnu} to find a random value for the mean GW frequency at the
detector. Since we have assumed a normalized GW spectral profile of the
source, $f(\nu)d\nu$, which has a mean frequency of $\bar{\nu}$, we may
find the mean GW frequency at the detector given a single GW/GRB
burst. That is, by specifying both the mean frequency at the source and at
the detector, we have randomly selected the source redshift. Using this
redshift we can find the full spectrum at the detector by multiplying our
selected value for $S^{\prime}$ by the redshifted spectral profile,
$f^{\prime}_{\bar{\nu}}(\nu^{\prime})d\nu^{\prime}$.

\subsection{GW detection}
\subsubsection{General}
We may now discuss the detection of such a GW burst. We assume that the GW
burst bandwidth is much greater than that of the bar detector since the
later is typically of the order of $\sim 0.3{\rm ~Hz}$ (Tobar et
al. 1995). The cross section for a single bar resonance in response to GWs
in such a situation is given by (Paik \& Wagoner 1976)
\begin{equation}
\label{eq:MTW-sigma}
\Sigma_{\nu^{\prime}_0} = \frac{32}{15}\frac{M}{\pi}\left(\frac{G}{c}\right)\left(\frac{v}{c}\right)^2\left\{ 1+O\left[ \left( \frac{R}{L}\right)^2 \right] \right\}\, ,
\end{equation}
where $M$ is the mass of the bar, $v$ is the speed of sound in the bar, $R$
is the radius of the physical cross section and $L$ is the length of the
bar. Eq. \ref{eq:MTW-sigma} assumes that, on the average, GW bursts are
incident isotropically on the detector with random polarisations. In most
bar detectors, $R \ll L$ and so the second order terms are neglected. We
can then find the amplitude of vibration (in strain) that the absorption of
GWs causes in the bar (Misner, Thorne \& Wheeler 1973, p. 1039),
\begin{equation}
\label{eq:MTW-h}
h = \frac{1}{L}\left(\frac{S^{\prime}\Sigma_{\nu^{\prime}_0}f^{\prime}_{\bar{\nu}}(\nu^{\prime})d\nu^{\prime}|_{\nu^{\prime}_0}}{2\pi^2(\nu_0^{\prime})^2 M }\right)^{1/2}\, .
\end{equation}

\subsubsection{The Niobium bar detector at UWA}
The discussion above has remained general for most cylindrical bar
detectors. However, for the remainder of the analysis, we take the specific
example of the Niobium bar detector at the University of Western Australia,
NIOBE, since we wish to model the detection algorithms in order to estimate
the signal seen at the outputs. The results will carry a factor of $\sim
5$--$10$ uncertainty and so they will remain representative of most bar
detectors. In particular, we consider the simplest algorithm to analyse GW
data streams, the zero order prediction (ZOP) technique. Here we summarize
some main points. For details of the detection methods used for NIOBE and
for descriptions of the ZOP algorithm, see Tobar et al. (1995), Dhurandhar,
Blair \& Costa (1996) and Heng et al. (1996).

GW data are sampled at $0.1{\rm ~s}$ intervals and passed through an
optimal low-pass filter before being decimated to $1{\rm ~Hz}$
samples. Correction factors for the low-pass procedure are included and the
results are scaled to noise temperature or strain. For an $\sim 1{\rm ~ms}$
burst of GWs, the rise time for the amplitude of vibration in the detector
is also $\sim 1{\rm ~ms}$. However, due to the high $Q$-factor, the
amplitude decays exponentially over a time scale of $\sim 200{\rm ~s}$. The
fact that the data are decimated means that the signal seen after the data
are processed is essentially the time derivative of the data. That is, if
we model the decay in strain amplitude by
\begin{equation}
\label{eq:h(t)}
h(t) = \sqrt{2}\, h \left[ \exp\left(-t\frac{dt}{T_{\rm mode}}\right) - \exp\left(-t\frac{dt}{T_{\rm int}}\right) \right]
\end{equation} 
then we may obtain an estimate of the signal after data processing using
\begin{eqnarray}
\label{eq:hbar}
\bar{h}(t) = \sqrt{2}\, h \left| \frac{dt}{T_{\rm mode}}\exp\left(-t\frac{dt}{T_{\rm mode}}\right) \right.-\nonumber \\ \left.\frac{dt}{T_{\rm int}}\exp\left(-t\frac{dt}{T_{\rm int}}\right) \right|\, ,
\end{eqnarray}
where $t$ is a dimensionless measure of the time after the GW burst that
takes into account the sampling time in the decimated data. $T_{\rm mode}$
is the ring down time for the mode of oscillation in the bar, $T_{\rm int}$
is the integration time for data samples and $dt$ is the time step for data
sampling. In our case $dt = T_{\rm int}$ and $T_{\rm mode} \sim 200{\rm
~s}$. This analysis does not take account of the noise floor of the
detector which will completely swamp such a single GW burst signal. Also,
we do not take account of any attenuation of the signal resulting from the
low-pass filtering; we assume that the processing algorithm corrects for
this.

\subsection{Co-addition Simulation}
We are now in a position to set up a Monte Carlo simulation to estimate the
signal in Eq. \ref{eq:hbar}. We must choose a model for the source so as to
estimate the GW spectral profile and the mean total GW energy output per
burst. The formalism above implicitly assumes that all GW bursts last for
less than $\sim 1{\rm ~s}$ and this must be reflected in the model of the
GW emission. We must also assume a redshift distribution of sources. The
models considered here will be discussed in Section 3. Given one of the set
of models, however, we may find a value for
$S^{\prime}f^{\prime}_{\bar{\nu}}(\nu^{\prime})d\nu^{\prime}|_{\nu^{\prime}_0}$
and use Eqs. \ref{eq:MTW-sigma} and \ref{eq:MTW-h} to find the signal after
the data processing, Eq. \ref{eq:hbar}. By using the cumulative
distributions, Eqs. \ref{eq:lumdist} and \ref{eq:cumnu}, we can build up a
simulation of many bursts and find their collective effect on the detector.

Throughout this work we assume, as a worst case estimate, that GRB times
are {\it uniformly} distributed over $10{\rm ~s}$ after the emission of the
corresponding GW bursts (Kochanek \& Piran 1993). However, we provide a
scaling of our results for a range of time lags in Section 4 (see Fig. 2)
since a less conservative view would put most GW bursts only $\sim 1{\rm
~s}$ before the GRB. The `GRB time' may be the trigger time in the BATSE
catalogue or it may be the time at which the gamma ray flux sees a
significant increase over threshold. The difference between such
definitions could be, at most, $\sim 10{\rm ~s}$. This issue is complicated
by the irregular and complex intensity profiles of GRBs (eg. Fishman \&
Meegan 1995). Here, for purposes of simulation, we assume that this problem
is resolved so that all the `GRB times' are aligned.

Finally, we note that the treatment above is valid only for a bar detector
with a single resonance band. However, as our calculations are aimed at
simulating the effect on NIOBE, which has two resonant frequencies, the
plus mode at $\nu_{0+}^{\prime} \sim 713{\rm ~Hz}$ and the minus mode at
$\nu^{\prime}_{0-} \sim 694.5{\rm ~Hz}$, we make the following adjustment
to the formalism. NIOBE acts approximately as a coupled two mass
oscillator. As a rough approximation, we may therefore model the detector
as having two resonance bands with cross sections given by
Eq. \ref{eq:MTW-h} centred at the relevant resonant frequencies.

\section{Models}
In this section we consider the parameters associated with various GW/GRB
models. Most of the resulting consistent sets of parameters describe binary
coalescence models as these seem to be the most likely GW/GRB source if
indeed the GW/GRB assumption is correct (Kochanek \& Piran 1993). The
remaining sets are, therefore, more speculative and provide useful
comparisons with the binary coalescence models in view of the final results
in Section 4.

In order to define a specific GW/GRB model we must specify the following.
\begin{enumerate}
\item {\bf Burst Mechanism:} Among the competing mechanisms for GRBs are NS
collision and coalescence (e.g. Piran 1991; Kochanek \& Piran 1993; Totani
1997; Ruffert \& Janka 1998) and other more general coalescence scenarios,
such as BH--NS (eg. Lee \& Klu\'{z}niak 1998) and white dwarf--BH
coalescences (eg. Fryer et al. 1999). These mechanisms all imply that the
source space density should follow (if somewhat lagged in time) the massive
star formation rate (eg. Totani 1997; Krumholz, Thorsett \& Harrison
1998). Other authors prefer to assume source space densities that are
similar to the quasar redshift distribution (eg. Horak, Emslie \& Hartmann
1995) without venturing as to what the mechanism for GRBs might be.

\item {\bf Spectral profile, $f(\nu)d\nu$:} For close binary systems, the
emission of GWs leads to energy loss and in-spiral. The GW frequency
therefore follows a `chirrup' pattern, monotonically increasing with
time. The Newtonian approximation for the in-spiral is a good one up until
the few rotation periods before coalescence when post-Newtonian corrections
become important. The innermost stable orbit marks the point at which GW
emission begins to decrease because of the increase in the infall rate of
the component stars. The innermost stable orbit occurs at an orbital
separation of $r \sim 6M$ (Kidder, Will \& Wiseman 1992, setting $G = c =
1$) and so Cutler \& Flanagan (1994) propose that the GW emission will shut
off at roughly $\bar{\nu} = (6^{3/2}\pi M)^{-1}$ for $M$ the total mass of
the binary system. The important parameters to obtain here are the mean
frequency of GW emission at the source and the spectral profile. For a
NS--NS system, the mass may lie in the range $M = (2.70 \pm 0.08)M_{\sun}$
(Thorsett \& Chakrabarty 1999). We therefore find that the frequency of
maximum emission is $\bar{\nu} \sim 1600{\rm ~Hz}$ and that the width of
the distribution of these frequencies is $\sim 100{\rm ~Hz}$. Assuming that
the `chirrup' is very steep near the coalescence and that the frequency of
emission is within $\sim 500{\rm ~Hz}$ of $\bar{\nu}$ for $\ll 1{\rm ~s}$
(eg. Allen 1997; Allen 1999) , we may assume, as we ultimately wish to
co-add signals at the detector, that the spectral profile at the source for
NS--NS coalescence has a width of $\sigma \sim 500{\rm ~Hz}$.

The issue of the actual value of $\bar{\nu}$ is of more importance here
since we are dealing with such a low bandwidth detector. However, it is a
poorly known quantity due to the slow convergence of post Newtonian
corrections to the estimate above. Using the results of post Newtonian
calculations of the final moments of in-spiral (Finn \& Chernoff 1993;
Cutler \& Flanagan 1994; Blanchet 1996) we find values in the range $\sim
1200$ to $\sim 1500{\rm ~Hz}$. We also note that estimates by Kidder, Will
\& Wiseman (1992, 1993), Damour, Iyer \& Sathyaprakash (1998) and Allen et
al. (1999) suggest that $\bar{\nu} = 1420,~2042.6{\rm ~and~} 1822{\rm ~Hz}$
respectively for $M = 2.8M_{\sun}$. It is therefore desirable to assume a
range of frequencies for each model of binary coalescence.

We also note that, for coalescence scenarios involving a BH, the higher
mass of the BH relative to that of a NS reduces the GW frequency. It is
therefore likely that GW emission from such systems will result in a low
flux at the detector resonant frequency. Consequently, we do not consider
such systems in our calculations.

With regards to NS collisions, little information can be gained as to the
general spectral profile of the source since this depends on the separation
of the path trajectories. Therefore, we do not consider collisions in our
calculations.

Finally, assuming that the GW/GRBs are associated with the quasar redshift
distribution gives no information on the distribution of frequencies at the
source; the GW mechanism has not been specified. We will therefore assume a
broad band emission in such an instance so as to compare with results from
the NS coalescence models. We will also consider a range of mean redshifts,
$z_{\rm m}$, in such models.

\item {\bf Mean total emitted GW energy, $F$:} For NS coalescences, this is
known to reasonable accuracy, but clearly, for our quasar model, we have no
information on this parameter. Let us therefore remove $F$ from
consideration in this instance by assuming that it is the same as the range
we will assume for NS coalescences. Various authors suggest that $F$ lies
in the range $10^{51}$--$10^{53}{\rm ~ergs}$ (eg. Rasio \& Shapiro 1992;
Centrella \& McMillan 1993; Oohara \& Nakamura 1995; Lee \& Klu\'{z}niak
1998) based on both numerical and approximate theoretical grounds. We
therefore provide calculations for this order of magnitude.

\item {\bf Number of bursts, N:} From the formalism in Section 2, we see
that if the number of bursts is large then the sum of all $h$ for all
bursts will scale approximately linearly with N. Therefore, throughout our
calculations we will assume that $N = 1000$ as a representative value.  We
compare this with the number of bursts in the BATSE 4Br catalogue, $N_{\sc
batse} = 1637$, and note that over $2000$ burst times have been
recorded. If the co-addition of GW data were to be performed, the number of
bursts available would be restricted by the fact that most bursts do not
have a simple intensity profile over the BATSE integration time. That is,
it may be difficult to assign a burst time to a given GRB due to the
complex nature of the profile. We therefore keep $N = 1000$ as an estimate
of the number of useful bursts for the current purpose.

In any co-addition of ``real'' data, the SNR will rise in proportion to
$\sqrt{N}$ for $N$ the number of data streams to be co-added. The noise is
co-added together with the signal; the signal rising in proportion to $N$
and the noise in proportion to $\sqrt{N}$.  However, our co-addition of
simulated data does not include the noise characteristics of the bar. If we
are to compare our final results with the sensitivity of the detector for a
{\it single} burst, the comparison must be between our noiseless results
and $\sqrt{N}$ times the typical noise at the resonance. We make this
comparison in Section 4.
\end{enumerate}

We have therefore narrowed our view of models of GW/GRBs to two main sets
of parameters. The first is the set associated with the assumption that
GRBs are due to coalescence of binary NS systems. We shall call this the
NS--NS model and take those parameters discussed above for it. We assume a
star formation redshift distribution that follows that of Madau, Pozzetti
\& Dickinson (1998) (hereafter MPD) and compare this with that given by
Pascarelle, Lanzetta \& Fern\'{a}ndez-Soto (1998) (hereafter PLFS). The
later compares well with the findings of Totani (1997) that the star
formation rate needs to be a factor of $5$--$10$ higher at redshifts $z
\gsim 2$ for the NS--NS scenario to be consistent with the BATSE data.

However, as a useful comparison, we parametrize the somewhat undefined
quasar model by assuming a Gaussian redshift distribution, centred on a
range of redshifts, $1.5 < z_{\rm m} < 3.0$ with a constant width, $\sigma =
0.45$ (Horack, Emslie and Hartmann 1995). The range in $z_{\rm m}$ may be
interpreted as a time lag factor (positive or negative) due to the specific
(unknown) mechanism of GW/GRBs. We also include such a time lag effect in
the star formation redshift distributions due to the length of time needed
for complete in-spiral of close NS--NS systems. We vary this time lag in
redshift space from $\sim 0.8$ to $0$ by allowing the redshift at which the
star formation rate first reaches a maximum, $z_{\rm p}$, to vary. We illustrate
some typical redshift distributions in Fig. 1.
\begin{figure}
\label{fig:zdist}
\centerline{\psfig{file=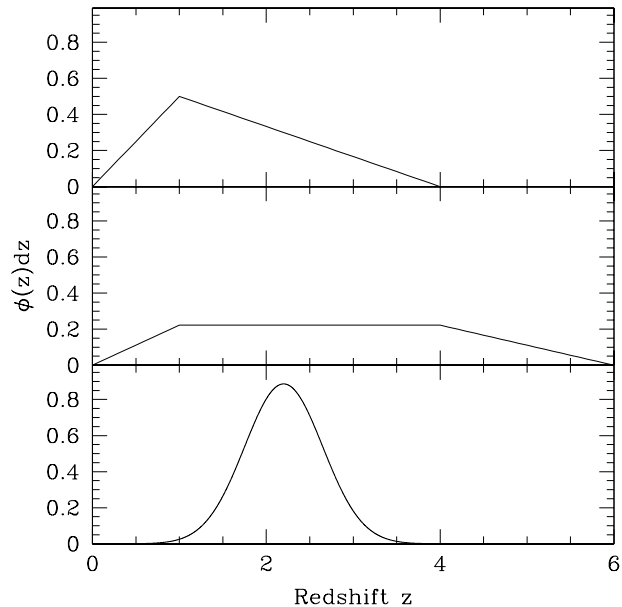,width=8.5cm}}
\caption{Comparison of three typical redshift distributions. The top
panel shows a MPD-like distribution and the middle panel shows a rough
fit to the redshift distribution in PLFS. The bottom panel shows the
quasar-like distribution used as a comparative example in our
calculations. All distributions are normalized to unity.}
\end{figure} 

\section{Results and Discussion}
For a specific model defined in the previous section, with a given mean
source frame frequency, $\bar{\nu}$, and a specific redshift distribution,
we obtain results typified by those shown in Fig. 2. Of note is the
distribution of $h$ shown by the points. These are co-added to give a
generally increasing signal, $\bar{h}(t)$, up until the common GRB time. At
times $t > 0$, the collective derivative patterns (Eq. \ref{eq:hbar}) of
all the events cause the decay of the signal.  The inset of Fig. 2 shows
the scaling for different values of the time lag between the GW burst and
the GRB. Although we have calculated this scaling with the same model
parameters as those described in the caption to Fig. 2, the scaling is
model independent and can be applied to any of our subsequent results
(Figs. $3$--$6$).
\begin{figure}
\label{fig:htime}
\centerline{\psfig{file=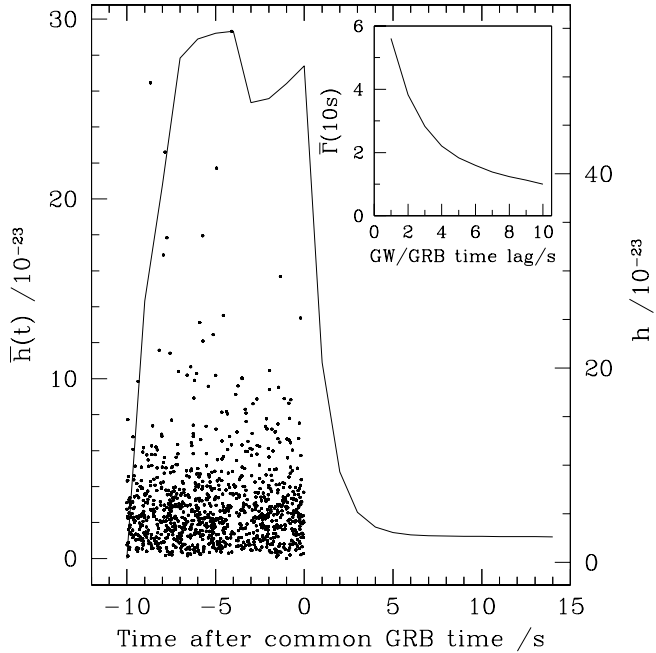,width=9.2cm}}
\caption{Demonstration of the co-addition of 1000 GW bursts. We used the
NS--NS scenario with the star formation redshift distribution of MPD (top
panel of Fig. 1) with $\bar{\nu} = 1600{\rm ~Hz}$ and $F = 10^{52}{\rm
~ergs}$. The plot shows each burst as a point on the $h$ scale and the
integrated signal seen after data processing is shown as the solid line. We
have assumed that burst times are distributed uniformly in time for the
$10{\rm ~s}$ before the common GRB time. The inset shows the maximum values
of $\bar{h}(t)$ for a range of time lag values relative to that for the
$10{\rm ~s}$ lag, $\Gamma(10{\rm s})$. For clarity, we have averaged the
maximum strain over 10 simulations to obtain $\bar{\Gamma}(10{\rm s})$.}
\end{figure}

Due to the fast decay time ($\sim 1{\rm ~s}$) of the signal output from the
ZOP algorithm (Eq. \ref{eq:hbar}), the time lag parameter is clearly an
important one. However, in a real data co-addition, the separation between
the GW burst time and the common GRB time is also subject to other
uncertainties as described in Section 2. We therefore prefer to assume that
these uncertainties limit us to an {\it effective} maximum difference
between the GW and GRB times of $10{\rm ~s}$ and we use this value in
calculating all subsequent results.

So as to give information for all models as functions of $\bar{\nu}$ and
the variable redshift parameter ($z_{\rm p}$ in the case of the NS--NS models and
$z_{\rm m}$ in the case of the quasar model), our results are presented as
three-surfaces in parameter space in Figs. $3$--$6$. Fig. 3 shows the
results for the NS--NS model with a MPD redshift distribution. The vertical
axis is the peak measured strain in the detector, ${\rm max}(\bar{h}(t))$,
in units of $10^{-23}$. We note the sharp rise in signal with increasing
mean source frequency and the broad leveling off for frequencies as high as
$\sim 1800$--$2200{\rm ~Hz}$. We also see a general increase in peak signal
with decreasing peak redshift which becomes less important at high source
frequency.

The noise on the surface is due to the fact that we only use 1000 bursts in
the calculations. Of course, we could have increased the number of bursts
to reduce the fractional fluctuation after the surfaces had been normalized
to 1000 bursts. However, the noise is illustrative of the expected
fluctuations in signal due to the various distributions (spacetime
distribution, distribution in frequency space etc.) that we assume the
sources follow. We do note, however, that the expected error in our
calculations may be as high as a factor of $5$--$10$ due to our basic
treatment in Sections 2 and 3 (Misner, Thorne \& Wheeler 1973,
pp. $1004$--$1044$). That is, we have not considered such parameters as the
bandwidth of the detector, aberrations due to coupling of the displacement
sensors to the bar etc. For this reason, we shall compare our results with
the total noise in the NIOBE detector later in this section.
\begin{figure}
\centerline{\psfig{file=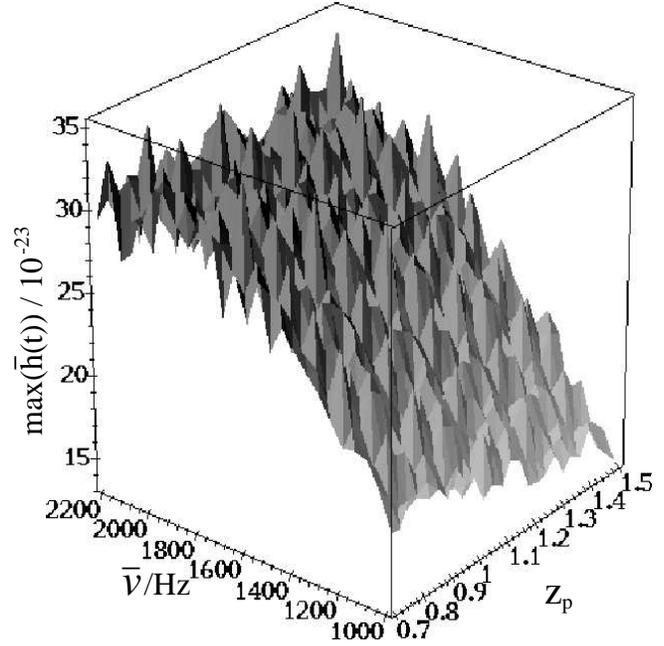,width=8.5cm}}
\caption{Three-surface showing the peak measured signal at the detector
outputs (after data processing), ${\rm max}(\bar{h}(t))$, as a function of
the peak redshift, $z_{\rm p}$, and the mean source frame frequency,
$\bar{\nu}$, for the NS--NS model with MPD redshift distribution. Model
parameters are described in Section 3. We used a mean total emitted GW
energy at the source of $F = 10^{52}{\rm ~ergs}$ but the vertical axis can
be scaled as $\sqrt{F}$ for generalization to other values.}
\end{figure}

In Fig. 4 we show results for a similar set of parameters (see Section 3)
but we use the PLFS redshift distribution. We note the immediate
qualitative difference in that the results show less variation in $z_p$
space and that there is a linear increase in the signal with increasing
frequency. This is similar to the almost linear increase we see in Fig. 3
(for the MPD model) at low frequencies and is due to the fact that more
sources lie at higher redshift in the PLFS model. That is, the frequency of
the leveling off will be higher since the GWs, on the average, at an
average redshift in the PLFS model, are redshifted below the resonance band
in the detector. Quantitatively though, the PLFS model provides a lower
average flux in the resonance band, and we see that the signal is $\sim
1/2$ that given by the MPD model throughout the parameter space.
\begin{figure}
\centerline{\psfig{file=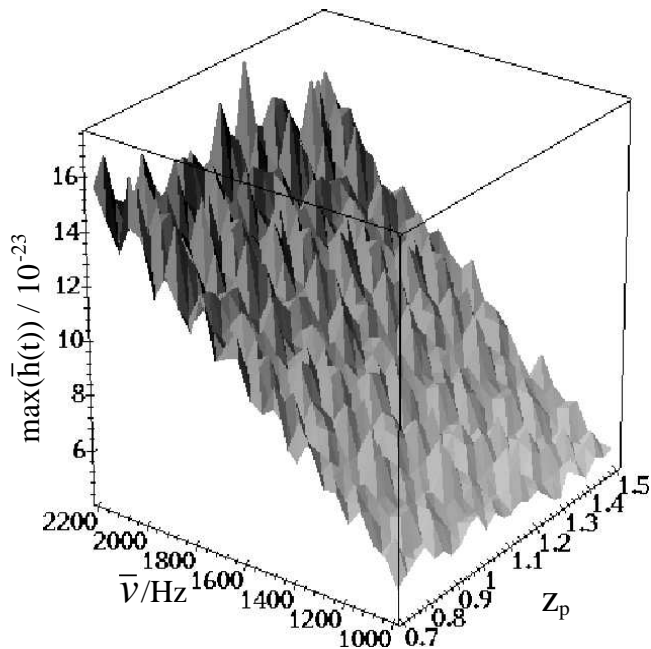,width=8.5cm}}
\caption{Three-surface showing the peak measured signal at the detector
outputs, ${\rm max}(\bar{h}(t))$, as a function of the peak redshift,
$z_p$, and the mean source frame frequency, $\bar{\nu}$, for the NS--NS
model with PLFS redshift distribution. Model parameters are described in
Section 3. We used a mean total emitted GW energy at the source of $F =
10^{52}{\rm ~ergs}$ but the vertical axis can be scaled as $\sqrt{F}$ for
generalization to other values.}
\end{figure}

Finally, in Figs. 5 and 6, we provide the results for the quasar model. For
Fig. 5 we used a uniform frequency distribution at the source with an order
of magnitude width. All other parameters are kept the same as those used
for the NS--NS models with the exception of the range of redshift parameter
used. Fig. 6 shows a similar calculation using a Gaussian frequency
distribution with the same width ($\sigma = 500{\rm ~Hz}$) as those used in
the NS--NS models. We see a much slower variation of the signal over the
parameter space, particularly at lower frequencies. Quantitatively, we see
that the quasar models have comparative signal level with the star
formation models and so we find that the signal seen at the bar detector is
reasonably insensitive to the redshift and source model.
\begin{figure}
\centerline{\psfig{file=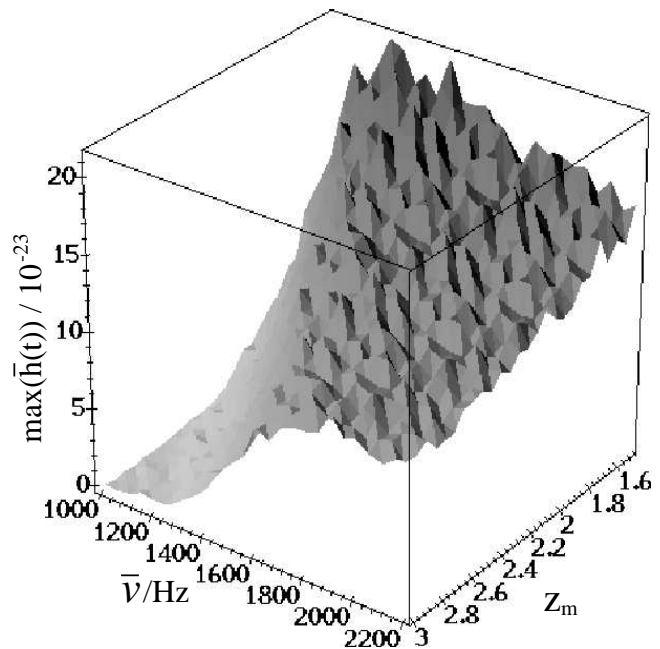,width=8.5cm}}
\caption{Three-surface showing the peak measured signal at the detector
outputs, ${\rm max}(\bar{h}(t))$, as a function of the mean redshift,
$z_{\rm m}$, and the mean source frame frequency, $\bar{\nu}$, for the quasar
model. Model parameters are described in the text and are set so that we
may compare these results with those in Figs. 3 and 4. In particular, we
used a mean total emitted GW energy at the source of $F = 10^{52}{\rm
~ergs}$ but the vertical axis can be scaled as $\sqrt{F}$ for
generalization to other values. However, here we have assumed a uniform
frequency profile at the source with a width of an order of magnitude.}
\end{figure}
\begin{figure}
\centerline{\psfig{file=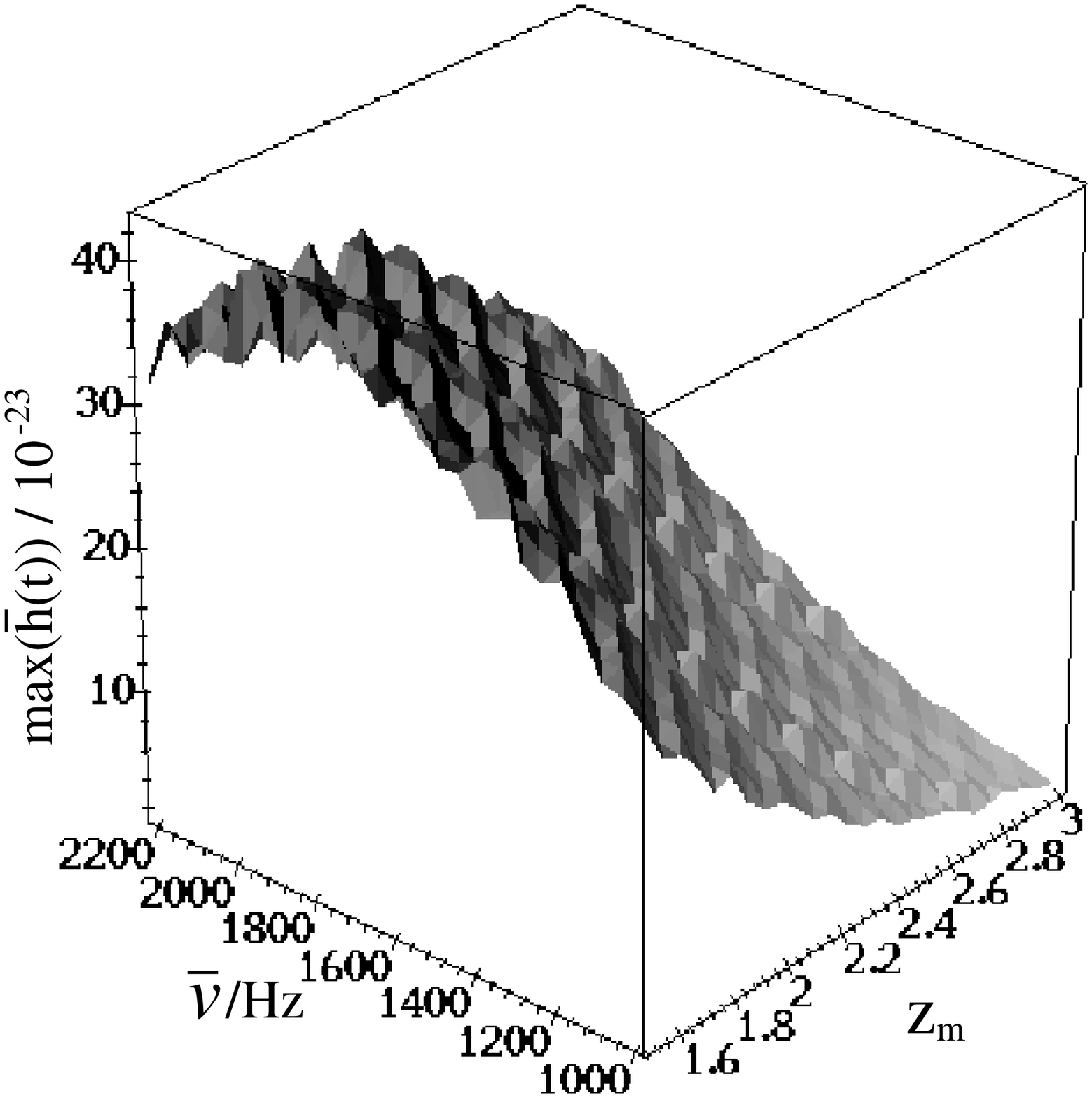,width=8.5cm}}
\caption{Three-surface showing the peak measured signal at the detector
outputs, ${\rm max}(\bar{h}(t))$, as a function of the mean redshift,
$z_{\rm m}$, and the mean source frame frequency, $\bar{\nu}$, for the quasar
model. Model parameters are described in the text and are set so that we
may compare these results with those in Figs. 3 and 4. Note that, in this
case, we also keep the frequency profile as a Gaussian with $\sigma =
500{\rm ~Hz}$ at the source. Also, we used a mean total emitted GW energy
at the source of $F = 10^{52}{\rm ~ergs}$ but the vertical axis can be
scaled as $\sqrt{F}$ for generalization to other values.}
\end{figure}

We may compare the results in Figs. $2$--$6$ with the sensitivity curves
provided by Tobar et al. (1995) for the NIOBE bar detector. The sensitivity
curves show that, within the two resonance bands, the root-mean-square
noise reaches a level of $\sim 2$ and $\sim 7 \times 10^{-20}$, for the
plus and minus mode respectively. For 1000 co-added bursts, the noise rises
by a factor of $\sqrt{1000}$; $\sim 6$ and $\sim 22 \times 10^{-19}$ in
each band respectively. However, the maximum signals calculated in
Figs. $3$--$6$ only range from $16$--$31 \times 10^{-23}$. That is, the
maximum expected signal-to-noise ratio after co-addition is $\sim 3 \times
10^{-4}$. Therefore, despite the potential gain in signal evident in the
co-addition results above, the NIOBE detector and other bar detectors, will
still be unable to detect the signal with 1000 GRB times and existing GW
data.

However, plans exist for improvements to be made to the NIOBE detector.  A
2-mode sapphire transducer is to replace the existing vibration sensors and
the cryogenic amplifier is to be significantly improved. In addition, the
(microwave source) pump oscillator driving the transducer is to be
improved. Tobar et al. (1998) report that these changes will decrease the
detector noise to $\sim 5 \times 10^{-21}$ in strain. A more important
parameter will be the increase in the bandwidth of the detector from $\sim
0.3{\rm ~Hz}$ in two bands to a single band with a width of $\sim 50{\rm
~Hz}$. We discuss the prospects for the improved detector in Section 5.

\section{Conclusions}
Using models of the GW/GRB source and the spacetime distribution of
sources, we have estimated the signal seen in the processed data of a
typical bar detector, NIOBE. Assuming a knowledge of the time of the GW
bursts (in this case, by assuming an association between GW and GRBs)
allows us to co-add simulated GW data from many bursts.  Figs. $3$--$6$
show the variation of the resultant signal with the main source parameter,
the mean GW frequency $\bar{\nu}$, and the main spacetime parameter, the
peak or mean redshift, $z_p$ or $z_{\rm m}$. These results assume
co-addition of 1000 measured burst events. The largest signal that can be
expected occurs within the NS--NS model with the MPD redshift
distribution. The peak in signal occurs over a large region of parameter
space covering mean rest frame GW frequencies $\bar{\nu} \gsim 1600{\rm
~Hz}$ and all realistic peak redshifts $z_p$. The maximum expected SNR is
$\sim 5 \times 10^{-4}$ in this case. Thus, comparing the signal with the
noise floor of NIOBE, we find that the co-addition, with the present
sensitivity of NIOBE and the present number of bursts, is insufficient to
allow detection by a method based on the formalism in Section 2.

However, the improvements to NIOBE will result in an order of magnitude
sensitivity gain. The bandwidth will also increase by a factor
$50.0/(2\times0.3) = 83$. Tobar \& Blair (1995) and Tobar (1997) report
that, to a good approximation, $h \propto \sqrt{\Delta f}$ for $\Delta f$
the bandwidth of the detector. That is, from Eq. \ref{eq:MTW-h}, a linear
change in the cross section of the bar, $\Sigma_{\nu_0^{\prime}}$,
corresponds to a change in $h \propto \sqrt{\Sigma_{\nu_0^{\prime}}}$. We
can therefore find an optimistic estimate of the number of bursts, $N_{\rm
b}$, required for us to achieve a SNR $\sim 1$,
\begin{equation}
\begin{array}{ll}
N_{\rm b}\sim&1000\left(\frac{1}{5\times10^{-4}}\right)^2\left(\frac{S_{\rm new}}{3\times10^{-20}}\right)^2\left(\frac{1}{\Gamma(10{\rm s})}\right)^2\\
 & \left(\frac{10^{52}{\rm~ergs}}{F}\right)\left(\frac{0.6{\rm ~Hz}}{\Delta
 f}\right)\, .\\
\end{array}
\end{equation}
The first term reflects the fact that we have achieved a SNR $\sim 5\times
10^{-4}$ using 1000 co-added bursts. The second term takes account of an
increase in sensitivity; $S_{\rm new}$ is the strain sensitivity in the
resonance band of the improved detector. The final terms incorporate
changes in the time lag factor introduced in Fig. 2, an increased average
energy output in GWs at the source and a change in the bandwidth of the
detector respectively. Thus, the number of GW/GRBs to be observed could be
as low as $\sim 8 000$--$10 000$ in the best case\footnote{That is, we use
$F \sim 10^{53}{\rm ~ergs}$ and a time lag of $2{\rm ~s}$ between GW and
GRBs. The lowest limit on the time lag is $1{\rm ~s}$ for NIOBE due to the
ZOP processing algorithm (the sampling time is $1{\rm ~s}$). We allow for a
further $1{\rm ~s}$ uncertainty introduced due to the difficulty in
defining the `GRB time' as discussed in Sections 2 and 4.}. Under this
assumption, detection of GWs by NIOBE could be found in $\sim 25$ years as
the GRB detection rate is $\sim 1$ per day. A disadvantage here is that the
accumulation of new observed bursts would start after the improvements are
made to NIOBE and the already available data would be of very limited use.

An additional possibility is that similar improvements be made to other
existing bar detectors. The data corresponding to GRB times from the five
detectors could then be co-added and the time for a possible detection
could reduce to only $25/\sqrt{5} \sim 10$ years assuming that all
detectors cover all GW/GRB events and that they all achieve similar
characteristics to the improved NIOBE detector.

Finn, Mohanty \& Romano (1999) have independently suggested a similar idea
to that presented in this article, regarding the future capabilities of the
LIGO detectors. They suggest that correlating the output from two LIGO
detectors previous to GRB times could result in a 95\% confidence detection
of an association between GRBs and GWs after the observation of $\sim 1000$
GRBs. If no signal is present then increased numbers of observed bursts
could place limits on the possible mechanisms for GRBs. Conceivably, a
detection could be possible within $4$--$5$ years of the beginning of full
LIGO operation. However, if the GW waveform is known (i.e. the GW/GRB
mechanism is determined to some degree), a detection of GWs may occur in
less time since matched filtering could be used to analyse the signal more
effectively.

The importance of the above results is clear. If improvements are made to
existing bar detectors, a detection of GW/GRB association may be possible
in a similar amount of time as for twin LIGO-like detectors using the
technique outlined in Sections 2 and 3. Such detection, whether by bar
detectors or by laser interferometers, may be the only observational
technique available to place constraints on GRB mechanisms aside from
measurements of fading optical GRB counterparts. This distinct advantage
comes with the loss of polarisation information and information about the
time delay between GW and gamma ray emission. These are, unfortunately,
important parameters for distinguishing gamma ray burster models (Kochanek
\& Piran 1993). Also, despite the fact that the results in Figs. $3$--$6$
are quite distinct, our ability to distinguish the different model
parameters is lost in the co-addition procedure. However, the importance of
these losses is not great when we remind ourselves that GWs have still not
been directly detected; both the lower cost improvements to existing bar
detectors and the high cost development of LIGO detectors are of great
importance to the study of GRBs.

\section*{Acknowledgments}
We would like to thank David Blair for initial discussions regarding this
work and for helpful comments about the manuscript and to Ralph Wijers and
Ken Lanzetta for a detailed discussion of neutron star physics. We are also
grateful to Michael Tobar and Eugene Ivanov for providing technical
information relating to NIOBE. We would also like to acknowledge useful
discussions with Alberto Fern\'{a}ndez-Soto with regards to suitable
redshift distributions and to Michael Ashley, Rocco Delillo and John
McMahon with regards to the status of GRB research. We also acknowledge a
helpful communication with Serge Droz and thank Melinda Taylor for computer
assistance.

\label{lastpage}
\end{document}